\documentclass{article}

 \PassOptionsToPackage{numbers, compress}{natbib}

 \usepackage[preprint]{nips_2018}

\usepackage[utf8]{inputenc} 
\usepackage[T1]{fontenc}    
\usepackage{hyperref}       
\usepackage{url}            
\usepackage{booktabs}       
\usepackage{amsfonts}       
\usepackage{nicefrac}       
\usepackage{microtype}      
\usepackage{amsmath}
\usepackage{amscd}
\usepackage{amsfonts}
\usepackage{amssymb}
\usepackage{amsthm}
\usepackage[dvips]{color}
\usepackage{epsfig}
\usepackage{graphics}
\usepackage[all]{xy}

 \newtheorem{lem}{Lemma}[section]
 
 \newtheorem{con}[lem]{Conjucture}
 \newtheorem{thm}[lem]{Theorem}
 
 \theoremstyle{definition}
 
 \theoremstyle{remark}

\title{Category coding with neural network application}

\author{
  Qizhi Zhang  \\
  \texttt{qizhi.zqz@alibaba-inc.com} \\
  \And
  Kuang-chih Lee \\
  \texttt{kuang-chih.lee@alibaba-inc.com}\\
  \AND
  Hongying Bao \\
  \texttt{hongying.bhy@alibaba-inc.com} \\
  \and 
  Yuan You \\
  \texttt{youyuan.yy@alibaba-inc.com} \\
  \and
  Dongbai Guo \\
  \texttt{dongbai.gdb@alibaba-inc.com}
}

\begin{document}

\maketitle

\begin{abstract}
In many applications of neural network, it is common to introduce huge amounts of input categorical features, as well as output labels. However, since the required network size should have rapid growth with 
respect to the dimensions of input and output space, there exists huge cost 
in both computation and memory resources. In this paper, we present a novel 
method called category coding (CC), where the design philosophy follows the principle of 
minimal collision to reduce the input and output dimension effectively. In addition,
 we introduce three types of category coding based on different Euclidean domains. Experimental results show that all three proposed methods outperform the existing state-of-the-art coding methods, such as standard cut-off and error-correct output coding (ECOC) methods.      

\end{abstract}

\section{Introduction}


In machine learning, many features are categorical, such as color, country, user id, item id, etc.
In the multi-class classification problem, the labels are categorical too.
The ordering relation doesn't exist among different values for these categories.
  Usually those categorical variables are represented by one-hot feature vectors. For example, red is encoded to 
  100, yellow to 010 and blue to 001. But if the number of categories are very huge, 
  for example the user id and item id in e-commerce applications, the one-hot 
  encoding scheme needs too many resources to compute classification results. 
  
  In the past years while SVM is widely used,
  ECOC (error-correct output coding) 
  method is proposed for handling huge numbers of output class labels. The idea of ECOC is to reduce a multi-class classification problem
  of huge number of classes to some two-class classification problems using 
  binary error-correct coding. But for the solution of handling huge number of input categorical features, the similar
  method doesn't exist, because the categories can not be separated by linear model, unless the one-hot encoding is used.

%

  In recent year, the deep neural network has great improvement in terms of performance and speed. The 
  coding method can be applied to deep neural network with some new beneficial 
  reform.

  In the classification problem, because the number of labels of a single 
  neural network need not to be binary, if we use a deep learning network as a 
base learner, it is not necessary to limit the code to be binary. In fact, there is a trade-off between
the class number of one base learner and the number of base learner used. According to information theory,
if we use  $p$ classes classifiers as basic classifiers to solve a classification problem of $N$-class,
we need at least $\lceil \log _p N \rceil $'s base learners. For example, if we need to solve a classifying problem of 1M's classes, 
and we use the binary classifier as base learners, we need at least 20 base learners. 
For some classical applications, for example, the CNN image 
classification, we need to build a CNN network for every binary classifier. It is huge cost for 
computation and memory resources. But if we combine different base learners with 1000 classes, we 
need at least 2 base learners. We know that the number of parameters in a Deep neural network is
usually big, hence using a small number of base learner benefits the reduction of the cost in 
computing and storage.   
  
  On the other hand, because the neural network has the ability of non-linear 
  representation, we can use the encoding for categorical features too. Can we 
  use classical error-correct coding for categorical features? We know that in machine learning,
  the sparsity is a basic rule to be satisfied, but the classical  error-correct coding does not
  satisfy the sparsity. Hence we need to design a new sparse coding scheme for this application.

In this paper, we give some new encoding method, they can be applied to both label encoding 
and feature encoding and give better performance than classical method. In section 2, we give the 
definition of category coding (CC) and
propose 3 classes of CC, namely Polynomial CC, Remainder CC and Gauss CC, which have good property.
In section 3 we discuss the application of CC in label encoding.
 In section 4, we discuss the application  
of CC in feature encoding. 
Our main tool is finite field theory and number theory, which can refer to \cite{ff} and \cite{NT}.

%
%
%
  
  

\vspace{-2mm}
    
\section{Category coding}
\vspace{-2mm}
For a $N$-class categorical feature or label,
we define a category coding (CC) as a  map 
\begin{displaymath} 
  \begin{array}{ccc}
   f: \mathbb{Z}/N\mathbb{Z} & \longrightarrow & \prod_{i=1}^r \mathbb{Z}/N_i\mathbb{Z} 
   \\
   x & \mapsto & (f_i(x))_i
  \end{array}
\end{displaymath} 
where each $f_i: \mathbb{Z}/N\mathbb{Z} \longrightarrow \mathbb{Z}/N_i\mathbb{Z}$ is called a ``site-position function''.
category coding, for  $i=1,2,...r$.

Generally, $N$ is a huge number, and $N_i$ are some numbers of middle size. 

We can
reduce a $N$-classes classification problem to $r$'s classification problems of 
middle size through a CC. 

We can also  use a $r$-hot $(\sum_{i=1}^nN_i)$-bit binary encoding instead of the 
one-hot encoding as the representation of the feature, i.e., use the composite of
the CC map $f$ and the nature embedding
\begin{displaymath}
  \begin{array}{ccl}
    \prod _{i=1} ^r \mathbb{Z}/N_i\mathbb{Z} & \longrightarrow & \prod_{i=1}^r \mathbb{F}_2^{N_i} = \mathbb{F}_2^{\sum_i N_i}\\
(x_i)_i & \mapsto &  (N_i \mbox{ bit one hot representation of } x_i )_i
  \end{array}
\end{displaymath}
to get a $r-$hot encoding. 

For a CC $f$, we call $
  \max_{x \neq y}  \sharp \{i=1, \cdots , r | f_i(x)=f_i(y)\}
$
the collision number of $f$, and denote $C(f)$. We have the following theorem.
\begin{thm}
  For a CC  $
    f: \mathbb{Z}/N\mathbb{Z}  \longrightarrow  
  \prod_{i=1}^r \mathbb{Z}/N_i\mathbb{Z} $
  , where $N_1 \leq N_2 \leq \cdots \leq 
  N_r$, we have $C(f)\geq \min\{i=1, \cdots r | N \leq \prod _{j=1} ^i N_j\}-1$.
\end{thm}
\textbf{Proof.} 
Let $k:=\min\{i=1, \cdots r | N \leq \prod _{j=1} ^i 
N_j\}$.
Suppose $C(f) < k-1$,
i.e
\begin{displaymath}
\max_{x \neq y}  \sharp \{i=1, \cdots , r | f_i(x)=f_i(y)\}<k-1
\end{displaymath}
Hence for any $x\neq y \in \mathbb{Z}/N\mathbb{Z}$,
there are at most $k-2$ same site-position value between $f(x)$ and $f(y)$.
Hence  $
\mathbb{Z}/N\mathbb{Z} \longrightarrow \prod_{i=1}^{k-1} \mathbb{Z}/N_i\mathbb{Z}
$
is an injection, and hence $N\leq \prod_{i=1}^{k-1}N_i$. It is a contradiction 
with the definition of $k$. \qed

If a CC satisfying $C(f)= \min\{i=1, \cdots r | N \leq \prod _{j=1} ^i N_j\}-1$, 
we call it has the \textbf{minimal collision} property.
In both usage of label encoding and feature encoding, we wish the code has 
minimal collision property.  

 We give
3 classes of CC, i.e, Polynomial CC, Remainder CC and Gauss CC, which satisfies the
minimal collision property.

\vspace{-2mm}
\subsection{Polynomial CC}
\vspace{-2mm}
For any prime number $p$, we can represent any non-negative integral number $x$ less than $p^k$ as the 
unique form
$x=x_0+x_1p+ \cdots + x_{k-1} p^{k-1} \quad (x_i \in \mathbb{Z}/p\mathbb{Z})$, which gives a bijection
$
  \mathbb{Z}/p^k\mathbb{Z} \longrightarrow \mathbb{F}_p ^k
$, 
where $\mathbb{F}_p$ is the Galois field (finite field) of $p$ elements.

For the classification problem of $N$-classes and any small positive integral number $k$ 
(for example, k=2, 3) and a small real number $\epsilon \in (0,1)$, 
we take a prime number in $[N^\frac{1}{k}, N^\frac{1}{k-\epsilon}]$ (According to the Prime Number Theorem ( \cite{Riemann}, \cite{Prime_Number_Theorem}), 
there are about $\frac{ k (N^\frac{1}{k-\epsilon}-N^\frac{1}{k} )}{\log N}$ such prime numbers.)  
, and get a injection
$
  \mathbb{Z}/N\mathbb{Z} \longrightarrow \mathbb{Z}/p^k\mathbb{Z} \longrightarrow \mathbb{F}_p ^k
$
by p-adic representation.

\begin{thm}
 For $r$'s different elements $x_1, x_2, \cdots, x_r$ in $\mathbb{F}_p$, the code 
 defined by the composite map $f$ of 
 the p-adic representation map 
and the map 
 \begin{displaymath}
   \begin{array}{rcl}
    \phi_1: \mathbb{F}^k_p & \longrightarrow & \mathbb{F}_p[x]_{deg<k}  \\
     (a_0, \cdots, a_{k-1} ) & \mapsto & a_0+a_1x+ \cdots a_{k-1} x^{k-1} 
   \end{array}
 \end{displaymath} 
 and the map
  \begin{displaymath}
   \begin{array}{rcl}
    \phi_2: \mathbb{F}_p[x]_{deg<k}  & \longrightarrow & \mathbb{F}_p^r  \\
     g(x) & \mapsto & (g(x_1), \cdots g(x_r) ) 
   \end{array}
 \end{displaymath} 
has the minimal collision property.
 \quad $\blacksquare$
  \end{thm}
\textbf{Proof.} 
We need proof that $C(\phi)\leq \min\{i=1, \cdots r | N \leq p^i\}-1$.
Because we know that $\min\{i=1, \cdots r | N \leq p^i\}=k$, hence we
need just prove $C(\phi)\leq k$, i.e for any $\alpha\neq \beta \in 
\mathbb{Z}/N\mathbb{Z}$, 
$
\sharp \{i=1, \cdots , r | f_i(\alpha)=f_i(\beta)\}\leq k-1
$.

Because the p-adic representation map and is an injection, and the 
map $\phi_1$ is a bijection, we need just to show that for any $g_1\neq g_2 \in 
\mathbb{F}_p[x]_{deg<k}$, 
$
\sharp \{i=1, \cdots , r | g_1(x_i)=g_2(x_i)\}\leq k-1
$.
Suppose there are $g_1\neq g_2 \in 
\mathbb{F}_p[x]_{deg<k}$ such that $
\sharp \{i=1, \cdots , r | g_1(x_i)=g_2(x_i)\}>k-1
$,
it means the polynomial $g_1-g_2 \in \mathbb{F}_p[x]$ of degree at most $k-1$ 
has at least $k$ roots, it is a contradiction with the Algebraic Basic Theorem on fields.
\qed 


\textbf{Remark.} The composite map of $\phi_1$ and $\phi_2$ in above theorem is known as 
Reed-Solomon code also
\cite{Reed_and_Solomon}. The Reed-Solomon code is a class
of non-binary MDS (maximal distinct separate) code \cite{Singleton}. 
MDS property is a excellent property
in error-corrected coding. But unfortunately, it has not 
find any nontrivial binary MDS code yet up to now. In fact, for some situation,
the fact that there are not any nontrivial binary MDS code is proved.  
(\cite{Guerrini_and_Sala} and Proposition 9.2 on p. 212 in \cite{Vermani} ). This
is an advantage of CC than ECOC in label encode also.
%
%
%
%
%
%
%
%
%
%
%
\vspace{-2mm}
\subsection{Remainder CC}
\vspace{-2mm}

For the original label's set $\mathbb{Z}/N\mathbb{Z}$, a small number k like 2, or 3, etc., and 
a small positive number
$\epsilon \in (0,1)$, select $r$'s  pairwise co-prime numbers
 $p_1, p_2, \cdots p_r$ in the domain $\left[N^\frac{1}{k}, N^\frac{1}{k-\epsilon}
\right)$. (According to the Prime Number Theorem ( \cite{Riemann}, \cite{Prime_Number_Theorem}), 
there are about $\frac{ k (N^\frac{1}{k-\epsilon}-N^\frac{1}{k} )}{\log N}$ 
 prime and hence pairwise co-prime numbers in this domain.)

We define the remainder CC as 
\begin{displaymath}
  \begin{array}{ccc}
    \mathbb{Z}/N\mathbb{Z} & \longrightarrow & \prod _{i=1}^n \mathbb{Z}/p_i\mathbb{Z} \\
    x & \mapsto & f_i(x) 
  \end{array}
\end{displaymath}
where $f_i(x)=x \mod p_i$, and $\{p_i\}$ is called its modules. Then we have the following proposition:
\begin{thm}
  \label{remainder_row_thm}
  The remainder CC has the minimal collision property.
\end{thm}

\textbf{Proof.} 
We need only to show that, for any $x \neq y \in \mathbb{Z}/N\mathbb{Z}$, there are at most $k-1$'s $i$ such, that 
  $f_i(x)=f_i(y)$.
  
Suppose there exist $k$'s different $i$ 
such, that $f_i(x)=f_i(y)$, we can suppose that
$
  f_i(x)=f_i(y)$ \quad  for $i=1,2, \cdots k$.
Then we have $x \equiv y \mod p_i$ for all $i=1,2, \cdots, k$.
Because $\{p_i\}$ are pairwise co-prime numbers, we have $x \equiv y \mod \prod_{i=1} ^k p_i$. But we 
know $x, y \in \{0, 1, \cdots N-1 \}$, which in $\{0, 1, \cdots \prod_{i=1} 
^k p_i-1\} $, hence $x=y$.
\qed

\vspace{-2mm}
\subsection{Gauss CC}
\vspace{-2mm}
We propose a CC based on the ring of Gauss integers \cite{Gauss} \cite{NT}, and so called Gauss CC.

We write the ring of Gauss integers as $\mathbb{Z}[\sqrt{-1}]:=\{a+b\sqrt{-1} \in \mathbb{C} | a,b \in 
\mathbb{Z}\}$. For a big integral number $N$, let $t$ is the minimal positive real 
number such that the number of Gauss integers in the closed disc $\overline{U_t(0)}$ 
is not less than $N$, i.e
$
   \sharp \overline{U_t(0)} \cap \mathbb{Z}[\sqrt{-1}] \geq N
$ and $\sharp \overline{U_{t-\epsilon}(0)} \cap \mathbb{Z}[\sqrt{-1}] < N
$ for any small $\epsilon>0$.
In general, we have $\sharp \overline{U_t(0)} \cap \mathbb{Z}[\sqrt{-1}]$ is about $\pi 
t^2$, hence we can get such $t$ about $\sqrt{N/\pi}$.

We can embed the original IDs to the Gauss integers in Gauss integers in the closed 
disc.
\begin{displaymath}
  \mathbb{Z}/N\mathbb{Z} \hookrightarrow \overline{U_t(0)} \cap \mathbb{Z}[\sqrt{-1}]
\end{displaymath}
Let $k$ be a small positive integral number, like 2,3, and $\epsilon'$ be a small positive real number.
 Let $p_1, p_2, \cdots, p_r$ 
be $r$ pairwise co-prime Gauss integral numbers satisfying
$
  |p_i|\in [(2t)^\frac{1}{k}, (2t)^\frac{1}{k-\epsilon'})  \quad \mbox{ for }i=1,2, \cdots, r.
$
We define the category mapping
\begin{displaymath}
  \begin{array}{ccc}
   \overline{U_t(0)} \cap \mathbb{Z}[\sqrt{-1}] & \longrightarrow & \prod _{i=1}^r 
  \mathbb{Z}[\sqrt{-1}]/(p_i)  \\
     z & \mapsto & (f_i(z))_i 
  \end{array}
\end{displaymath}
where $(p_i)$ means the principle ideal of $\mathbb{Z}[\sqrt{-1}]$ generated by $p_i$,  $f_i(z)=z \mod (p_i)$. 
$\{p_i\}$ is called the modules of this Gauss CC, and we have the following theorem.
 
\begin{thm}
  \label{gauss_row_thm}
  The Gauss CC has the minimal collision property.
\end{thm}

\textbf{Proof.} From the method to take $\{p_i\}$, we know $k=\min\{i=1, \cdots r | N \leq \prod _{j=1} ^i |\mathbb{Z}[\sqrt{-1}]/(p_j)| 
\}$.
Hence we need only to show that,     for any $x \neq y \in \overline{U_t(0)} \cap \mathbb{Z}[\sqrt{-1}]$, there are at most $k-1$'s $i$ 
such, that
  $f_i(x)=f_i(y)$.
  
Suppose there exist $k$'s different $i$ 
such, that $f_i(x)=f_i(y)$, we can suppose that
\begin{displaymath}
  f_i(x)=f_i(y) \quad \mbox{ for }i=1,2, \cdots k
\end{displaymath} 
Then we have $x - y \equiv 0 \mod (p_i)$ for all $i=1,2, \cdots, k$.

Because $\{p_i\}$ are pairwise co-prime Gauss integral numbers, hence $\{(p_i)\}$ are pairwise co-prime ideal of 
$\mathbb{Z}[\sqrt{-1}]$, and
we have $x - y \in \prod_{i=1} ^k (p_i)$. Hence
$
  \mathbf{Nm}(x-y) \in  \prod_{i=1} ^k (\mathbf{Nm}(p_i))\mathbb{Z}
$
i.e,
$
  |x-y|^2 \in  \prod_{i=1} ^k |p_i|^2\mathbb{Z}
$, 
and hence
$
  |x-y| \equiv 0 \mod  \prod_{i=1} ^k |p_i| 
$.
But we 
know $x, y \in \overline{U_t(0)} $, hence $|x-y|\leq 2t$.
On the other hand, we know $\prod_{i=1} ^k |p_i| >2t$, hence $|x-y|=0$, and 
hence $x=y$. \qed

\vspace{-2mm}

\section{Application for label encode}

\vspace{-2mm}

For a $N$-class classification problem,
we use a CC
\begin{displaymath} 
  \begin{array}{ccc}
f: \mathbb{Z}/N\mathbb{Z} & \longrightarrow & \prod_{i=1}^r \mathbb{Z}/N_i\mathbb{Z} \\
    z & \mapsto & (f_i(z))_i
  \end{array}
\end{displaymath} 
to reduce a $N$-classes classification problem to $r$'s classification problems of 
middle size through a LM. Suppose the training dataset is $\{x_k, y_k\}$, where $x_k$ is 
feature and $y_k$ is label, then we train a base learner on
the dataset $\{x_k,f_i(y_k)\}$ for every $i=1,2, \cdots r$. We call it the label 
encoding method.

A CC good for label encoding should satisfy the follow properties:

\textbf{Classes high separable}. For two different labels $y, \tilde{y}$, there should be as 
many as possible site-position functions $f_i$ such that $f_i(y)\neq f_i(\tilde{y})$.

\textbf{Base learners independence}. When $y$ are selected randomly uniformly from $\mathbb{Z}/N\mathbb{Z}$,
the mutual information of $f_i(y)$ and $f_j(y)$ approximate to 0 for $i\neq j$.

The property ``classes high separable'' ensures that for any two different classes, 
there are as many as possible base learners are trained to separate them.
 The property ``base learners independence'' ensures that the common part 
 of the information learned by any two different base learners is few.

\vspace{-1mm}

\textbf{Remark.}
These properties are the similar of the properties ``Row separable'' and ``Column separable'' of
 ECOC (\cite{Dietterich_and_Bakiri}) in non-binary situation. 
%
%
%

%
%
%
%

The minimal collision property ensure the CCs satisfy ``Class high 
separable'', we will show that they satisfy ``Base learner independence'' also.
\vspace{-2mm}
\subsection{Polynomial CC}

\vspace{-3mm}
%
%
%
%

We will prove that, the Polynomial CC satisfies the property ``Base learners independence'' also.



\begin{thm}
  \label{simplex_LM_Thm}
  If $u$ is a random variable with uniform distribution on 
$\mathbb{Z}/N\mathbb{Z}$, $y_i$ and $y_j$ are the i-site value and j-site value 
($i \neq j$) 
of the codeword of $u$ under the simplex LM described above, then the mutual information of 
$y_i$ and $y_j$ approach to $0$ when $N$ grows up. 
\end{thm}

\textbf{Proof.}

For any $u$ in $\mathbb{Z}/p^k\mathbb{Z}$, the i-th site value is
$
  y_i=u_0+u_1x_i+ \cdots u_{k-1}x_i^{k-1}  \quad \mod p
$, where $u_0, u_1, \cdots u_{k-1}$ are the coefficients of the p-adic 
representation of $u$. We denote this map by $g_i: \mathbb{Z}/p^k\mathbb{Z} \longrightarrow 
\mathbb{Z}/p\mathbb{Z}$.

Let $t=\lceil N/p \rceil$, consider the following commutative diagram: \quad
 \xymatrix{
    \mathbb{Z}/pt\mathbb{Z} \ar[r]\ar[d] ^{g_i}  &  \mathbb{Z}/pt\mathbb{Z} \ar[d]^{g_i}\\   
     \mathbb{Z}/p\mathbb{Z} \ar[r]  &   \mathbb{Z}/p\mathbb{Z} 
    }

The horizontal arrow in up line is defined by
$
  u_0+u_1p+ \cdots u_{k-1}p^{k-1} \mapsto (u_0+1 \mod p) +u_1p+ \cdots u_{k-1}p^{k-1}
$,
and the horizontal arrow in down line is defined by $y\mapsto (y+1 \mod p)$.
The horizontal arrows are bijections, which shows that the numbers of the pre-images 
in $\mathbb{Z}/pt\mathbb{Z}$ of every element in $\mathbb{Z}/p\mathbb{Z}$ are same and hence 
equal to $t$. 

On the other hand, we have the commutative diagram: \quad
 \xymatrix{
    \mathbb{Z}/p(t-1)\mathbb{Z} \ar[r]\ar[dr]   & \mathbb{Z}/N\mathbb{Z}  \ar[r]\ar[d] &  \mathbb{Z}/pt\mathbb{Z} \ar[dl]\\   
     & \mathbb{Z}/p\mathbb{Z} &   
    }

where the horizontal arrows are the natural embedding, and other arrows are the 
restriction of $g_i$.

But the number of pre-images in $\mathbb{Z}/pt\mathbb{Z}$
 of every element in $\mathbb{Z}/p\mathbb{Z}$ is $t$, and the same logic shows 
 that the number of pre-images in $\mathbb{Z}/p(t-1)\mathbb{Z}$
 of every element in $\mathbb{Z}/p\mathbb{Z}$ is $t-1$. Therefore the number of pre-images in 
 $\mathbb{Z}/N\mathbb{Z}$
 of every element in $\mathbb{Z}/p\mathbb{Z}$ is $t$ or $t-1$.
 
 Hence if $u$ is a random variable with uniformly distribution on $\mathbb{Z}/N\mathbb{Z}$, its 
 probability at every point in $\mathbb{Z}/N\mathbb{Z}$ is $1/N$, then the probability of $y_i$
 at every point in $\mathbb{Z}/p\mathbb{Z}$ are $\frac{t}{N}$ or $\frac{t-1}{N}$. 
 The same logic shows that the probability of $y_j$
 at every point in $\mathbb{Z}/p\mathbb{Z}$ are $\frac{t}{N}$ or 
 $\frac{t-1}{N}$.

Let $s=\lceil N/p^2 \rceil$, we 
 have the commutative diagram for any $(a,b) \in \mathbb{F}_p^2$:  
 \xymatrix{
    \mathbb{Z}/p^2s\mathbb{Z} \ar[r]\ar[d] ^{(g_i,g_j)}  &  \mathbb{Z}/p^2s\mathbb{Z} \ar[d]^{(g_i, g_j)}\\   
     \mathbb{Z}/p\mathbb{Z}\times \mathbb{Z}/p\mathbb{Z} \ar[r]  &   \mathbb{Z}/p\mathbb{Z} \times \mathbb{Z}/p\mathbb{Z} 
    }
 
where the up horizontal arrow is defined by
$
  u_0+u_1p+\cdots u_{k-1}p^{k-1} \mapsto (u_0+a \mod p)+ (u_1+b \mod p)p+\cdots  u_{k-1}p^{k-1}
  $,
and the down horizontal arrow is defined by
$
  (y_i,y_j) \mapsto (y_i+a+bx_i \mod p, y_j+a+bx_j \mod p)
$.
Both the horizontal arrows are bijections.

Because $x_i \neq x_j$ we know that when $(a,b)$ runs over all the pairs in 
$\mathbb{Z}/p\mathbb{Z} \times \mathbb{Z}/p\mathbb{Z}$  the down horizontal map 
maps $(0,0)$ to all the pairs in $\mathbb{Z}/p\mathbb{Z} \times 
\mathbb{Z}/p\mathbb{Z}$. Therefore all the number of pre-images in $\mathbb{Z}/p^2s\mathbb{Z}$ 
of any element in $\mathbb{Z}/p\mathbb{Z} \times 
\mathbb{Z}/p\mathbb{Z}$ are same, and hence equal to $s$.

A similar method shows that if $u$ is a random variable with uniformly distribution on $\mathbb{Z}/N\mathbb{Z}$, 
the joint probability of $(y_i, y_j)$
 at every point in
 $\mathbb{Z}/p\mathbb{Z} \times \mathbb{Z}/p\mathbb{Z}$ are $\frac{s}{N}$ 
 or $\frac{s-1}{N}$.

We know that the mutual information of 
$y_i$ and $y_j$ is 
$
 I(Y_i;Y_j)=\sum_{(y_i,y_j)\in \mathbb{Z}/p\mathbb{Z} \times \mathbb{Z}/p\mathbb{Z}}  
 p_{i,j}(y_i, y_j)\log \frac{ p_{i,j}(y_i, y_j)}{p_i(y_i)p_j(y_j)}
$.

a.) When $k=2$, i.e. $p< N \leq p^2$, we know $s=1$ and $p_{i,j}(y_i, y_j)=\frac{1}{N}$ on $N$'s
point in $\mathbb{Z}/p\mathbb{Z} \times \mathbb{Z}/p\mathbb{Z}$ and $0$ on 
other points. Hence we have
\begin{displaymath}
  \begin{array}{rl}
 I(Y_i;Y_j) & \leq \sum_{(y_i,y_j)\in \mathbb{Z}/p\mathbb{Z} \times \mathbb{Z}/p\mathbb{Z}}  
 p_{i,j}(y_i, y_j)\log \frac{ p_{i,j}(y_i, y_j)}{(\frac{t-1}{N})^2} 
 =N \times \frac{1}{N} \log \frac{ 1/N}{(\frac{t-1}{N})^2} 
  =2\log \frac{N}{t-1} -\log N  \\
 & \leq 2\log \frac{N}{N/p-1} -\log N 
  =2\log p -2\log (1-\frac{p}{N})-\log N 
   =2\log p+2O(\frac{p}{N})-\log N
 \end{array}
\end{displaymath}
However, $p \in [N^\frac{1}{2}, N^\frac{1}{2-\epsilon}]$ implies that $p=N^{\frac{1}{2}}(1+o(1))$, hence we have
\begin{displaymath}
 I(Y_i;Y_j) =\log N+ 2\log(1+o(1))+2O(N^{-\frac{1}{2}})-\log N  =o(1)  \rightarrow 0  \mbox{ as } N \rightarrow \infty
\end{displaymath}

 b.) When $k>2$, i.e. $ N>p^2$, we have
 \begin{displaymath}
   \begin{array}{rl}
      I(Y_i;Y_j) = & \sum_{(y_i,y_j)\in \mathbb{Z}/p\mathbb{Z} \times \mathbb{Z}/p\mathbb{Z}}  
 p_{i,j}(y_i, y_j)\log p_{i,j}(y_i, y_j)  - \sum_{(y_i,y_j)\in \mathbb{Z}/p\mathbb{Z} \times \mathbb{Z}/p\mathbb{Z}}  
 p_{i,j}(y_i, y_j)(\log p_i(y_i)+ \log p_j(y_j))          \\
  =& \sum_{(y_i,y_j)\in \mathbb{Z}/p\mathbb{Z} \times \mathbb{Z}/p\mathbb{Z}}  
 p_{i,j}(y_i, y_j)\log p_{i,j}(y_i, y_j)  - \sum_{y_i\in \mathbb{Z}/p\mathbb{Z}}  
 p_i(y_i)\log p_i(y_i)-  \sum_{y_j\in \mathbb{Z}/p\mathbb{Z}}  
 p_j(y_j)\log p_j(y_j)    \\
 \leq & p^2 \frac{s}{N} \log(\frac{s}{N})  -2p \frac{t-1}{N}\log \frac{t-1}{N}
   \end{array}
 \end{displaymath}
 Because $(s-1)p^2  <N \leq  sp^2$ and $(t-1)p    <N \leq  tp$,
we have 
 \begin{displaymath}
   \begin{array}{rl}
      I(Y_i;Y_j)     
    < & (1+\frac{p^2}{N}) \log (\frac{1}{p^2}+\frac{1}{N}) -2(1-\frac{p}{N}) 
    \log (\frac{1}{p}-\frac{1}{N}) 
  =  \log \frac{\frac{1}{p^2}+\frac{1}{N}}{(\frac{1}{p}-\frac{1}{N})^2}
      +\frac{p^2}{N}\log (\frac{1}{p^2}+\frac{1}{N})+2\frac{p}{N}\log (\frac{1}{p}-\frac{1}{N})  
      \\
 =& \log \frac{1+\frac{p^2}{N}}{ (1-\frac{p}{N})^2}
    +\frac{p^2}{N}   (\log (1+\frac{p^2}{N})-2\log p) +2\frac{p}{N}(\log (1-\frac{p}{N})-\log p)  
  < \log \frac{1+\frac{p^2}{N}}{ (1-\frac{p}{N})^2}
    +\frac{p^2}{N}   \log (1+\frac{p^2}{N})  \\
  =& O(\frac{p^2}{N})+O(\frac{p}{N})+ \frac{p^2}{N}O(\frac{p^2}{N}) 
  =  O(\frac{p^2}{N})
   \end{array}
 \end{displaymath}
 
However, $p\in [N^\frac{1}{k}, N^\frac{1}{k-\epsilon}]$ implies that  $p=N^{\frac{1}{k}}(1+o(1))$, hence we have
 \[
 I(Y_i;Y_j)=O(N^{\frac{2}{k}-1}) \rightarrow 0 \mbox{ as } N\rightarrow 
   \infty\]  \qed

\vspace{-4mm}

\subsection{Remainder CC and Gauss CC}
\vspace{-3mm}
%
%
%
%
%
%
%
%
%

The theorem \ref{remainder_row_thm}, \ref{gauss_row_thm} tells us that the Remainder CC and Gauss CC  satisfies the ``Classes high separable'' property. 
In fact, they satisfy the property ``Base learners independence'' also.
\begin{thm}
  Let  $f: \mathbb{Z}/N\mathbb{Z}  \longrightarrow  \prod _{i=1}^r 
  \mathbb{Z}/N_i\mathbb{Z}$ be a Remainder CC 
  , and
$x$ be uniformly randomly selected from $\mathbb{Z}/ N\mathbb{Z}$, we have that  
for any $i \neq j$,
  the mutual Information of $f_i(x)$ and $f_j(x)$ approximate 0.   
\end{thm}

\textbf{Proof.} 

Let $t_i:= \lceil \frac{N}{p_i} \rceil$ and  $s_{ij}= \lceil \frac{N}{p_ip_j} \rceil$ for every $i, j$. 
We have that the 
probabilities
of $f_i(x)$ at every point in $\mathbb{Z}/p_i\mathbb{Z}$ are $\frac{t_i}{N}$ or $\frac{t_i-1}{N}$ and
the  probabilities
of $(f_i(x), f_j(x))$ at every point in  $\mathbb{Z}/p_i\mathbb{Z} \times  \mathbb{Z}/p_j\mathbb{Z}$ 
are $\frac{s_{ij}}{N}$ or $\frac{s_{ij}-1}{N}$
by using the similar method in the proof of Theorem \ref{simplex_LM_Thm}.

We know that the mutual information of 
$y_i=f_i(x)$ and $y_j=f_j(x)$ is 
\begin{displaymath}
 I(Y_i;Y_j)=\sum_{(y_i,y_j)\in \mathbb{Z}/p\mathbb{Z} \times \mathbb{Z}/p\mathbb{Z}}  
 p_{i,j}(y_i, y_j)\log \frac{ p_{i,j}(y_i, y_j)}{p_i(y_i)p_j(y_j)}
\end{displaymath}

a.) When $k=2$, we have $N<p_ip_j$ and hence $s=1$ and $p_{i,j}(y_i, y_j)=\frac{1}{N}$ on $N$'s
point in $\mathbb{Z}/p\mathbb{Z} \times \mathbb{Z}/p\mathbb{Z}$ and $0$ on 
other points. Hence we have
\begin{displaymath}
  \begin{array}{rl}
 I(Y_i;Y_j) & \leq \sum_{(y_i,y_j)\in \mathbb{Z}/p\mathbb{Z} \times \mathbb{Z}/p\mathbb{Z}}  
 p_{i,j}(y_i, y_j)\log \frac{ p_{i,j}(y_i, y_j)}{\frac{(t_i-1)(t_j-1)}{N^2}} \\
 & =N \times \frac{1}{N} \log \frac{ 1/N}{\frac{(t_i-1)(t_j-1)}{N^2}} \\
& =\log N-\log(t_i-1) -\log (t_j-1)  \\
& < \log N -\log(\frac{N}{p_i}-1) -\log(\frac{N}{p_j}-1) \\
& \leq \log N - 2 \log( \frac{N^\frac{1}{2}}{N^{2-\epsilon} }-1) \\
& = -2\log(\frac{1}{N^{2-\epsilon}}-N^{-\frac{1}{2}}) \\
& \rightarrow 0 \quad \mbox{ as } N\rightarrow \infty
 \end{array}
\end{displaymath}

b.) When $k\geq 3$, we have $p_ip_j<N^\frac{2}{k-\epsilon}<N$, and

 \begin{displaymath}
   \begin{array}{rl}
     & I(Y_i;Y_j)     \\
     = & \sum_{(y_i,y_j)\in \mathbb{Z}/p_i\mathbb{Z} \times \mathbb{Z}/p_j\mathbb{Z}}  
 p_{i,j}(y_i, y_j)\log p_{i,j}(y_i, y_j)  \\
 & - \sum_{(y_i,y_j)\in \mathbb{Z}/p_i\mathbb{Z} \times \mathbb{Z}/p_j\mathbb{Z}}  
 p_{i,j}(y_i, y_j)(\log p_i(y_i)+ \log p_j(y_j))          \\
  =& \sum_{(y_i,y_j)\in \mathbb{Z}/p_i\mathbb{Z} \times \mathbb{Z}/p_j\mathbb{Z}}  
 p_{i,j}(y_i, y_j)\log p_{i,j}(y_i, y_j)  \\
 & - \sum_{y_i\in \mathbb{Z}/p_i\mathbb{Z}}  
 p_i(y_i)\log p_i(y_i)-  \sum_{y_j\in \mathbb{Z}/p_j\mathbb{Z}}  
 p_j(y_j)\log p_j(y_j)    \\
 \leq & p_i p_j \frac{s_{ij}}{N} \log(\frac{s_{ij}}{N})  -p_i \frac{t_i-1}{N}\log \frac{t_i-1}{N} 
 -p_j \frac{t_j-1}{N}\log \frac{t_j-1}{N}   
   \end{array}
 \end{displaymath}
 Because

 \begin{displaymath}
   \begin{array}{rcl}
     (s_{ij}-1)p_i p_j & <N \leq & s_{ij}p_ip_j \\
     (t_i-1)p_i   & <N \leq & t_ip_i \\
     (t_j-1)p_j   & <N \leq & t_j p_j
   \end{array}
 \end{displaymath}

We have 
 \begin{displaymath}
   \begin{array}{rl}
     & I(Y_i;Y_j)     \\
    < & (1+\frac{p_ip_j}{N}) \log (\frac{1}{p_ip_j}+\frac{1}{N}) \\
    &   -(1-\frac{p_i}{N}) 
    \log (\frac{1}{p_i}-\frac{1}{N})  -(1-\frac{p_j}{N}) 
    \log (\frac{1}{p_j}-\frac{1}{N}) \\
  = & \log \frac{\frac{1}{p_ip_j}+\frac{1}{N}}{(\frac{1}{p_i}-\frac{1}{N})(\frac{1}{p_j}-\frac{1}{N})}
  +\frac{p_ip_j}{N} \log (\frac{1}{p_ip_j}+\frac{1}{N}) \\
&  +\frac{p_i}{N}  \log (\frac{1}{p_i}-\frac{1}{N})
  +\frac{p_j}{N} \log (\frac{1}{p_j}-\frac{1}{N})  \\
 \leq & 
 \log(1+\frac{p_ip_j}{N})-\log(1-(\frac{1}{p_i}+\frac{1}{p_j})\frac{p_ip_j}{N}+\frac{1}{N^2}) 
 \\
 \leq & 
 \log(1+\frac{p_ip_j}{N})-\log(1-(\frac{1}{p_i}+\frac{1}{p_j})\frac{p_ip_j}{N}) 
 \\
 = & O(\frac{p_ip_j}{N})+O((\frac{1}{p_i}+\frac{1}{p_j})\frac{p_ip_j}{N}) \\
 = & O(\frac{p_ip_j}{N})\\
 = & O(N^{\frac{2}{k-\epsilon}}-1) \\
 = & O(N^\frac{2+\epsilon-k}{k-\epsilon}) 
  \rightarrow 0 \quad \mbox{as} \quad  N\rightarrow \infty     
   \end{array}
 \end{displaymath}

$\blacksquare$

This theorem tells us that, the Remainder CC  satisfies the property ``Base learners 
independence''.

Similarly, we have
\begin{thm}
  Let  $f: \mathbb{Z}/N\mathbb{Z}  \longrightarrow  \prod _{i=1}^r 
  \mathbb{Z}/N_i\mathbb{Z}$ be a Gauss CC, and
$x$ be uniformly randomly selected from $\mathbb{Z}/ N\mathbb{Z}$, we have that  
for any $i \neq j$,
  the mutual Information of $f_i(x)$ and $f_j(x)$ approximate 0.   
\end{thm} \qed

This theorem tells us that, the Gauss CC  satisfies the property ``Base learners independence'' 
also.

\vspace{-2mm}
\subsection{Decode Algorithm}
\vspace{-2mm}
Suppose we used the LM
$
f_i: \mathbb{Z}/N\mathbb{Z} \longrightarrow \mathbb{Z}/N_i\mathbb{Z} \quad (i=1,2,...n) 
$
to reduce a classification problem of class number $N$ to the classification problems of 
class number $N_i$'s, and trained $n$ base learner for every $f_i$, the output 
of every base learner $i$ is a distribution $P_i$ 
on $\mathbb{Z}/N_i\mathbb{Z}$. Now, for a input feature data,
how we collect the output $\{P_i : i=1,2,\cdots, n\}$ of every
base learner to get the predict label?
    
In this paper, we search the $x\in \mathbb{Z}/N\mathbb{Z}$ such that $\sum_i \log P_i(f_i(x))$ 
is maximal, and let such $x$ be the decoded label. (In fact, $\sum_i \log P_i(f_i(a))=
-\sum_i KL(f_{i\star} \delta(x-a) ||  P_i)$
,  where $\delta(x-a)$ is the Delta distribution at $a\in \mathbb{Z}/N\mathbb{Z}$, and $f_{i\star} \delta(x-a)$ is 
the marginal distribution of $\delta(x-a)$ induced by $f_i$.)

\vspace{-2mm}

\subsection{Numeric Experiments}

\vspace{-2mm}
We use the Inception V3 network and LM on the  dataset ``CJK characters''. CJK is a collective term for the Chinese, Japanese, and Korean languages, all of which use 
Chinese characters and derivatives (collectively, CJK characters) in their writing systems. 
The data set ``CJK characters'' is the grey-level image of size 139x139 of 20901 CJK characters (0x4e00 $\sim$ 0x9fa5)
in 8 fonts.

%

 We use 7
fonts as the train set, and other one font as the test set.  We use inception v3 
network as base learner, and train the networks using batch size=128 and 100 batch per an epoch.

We use three CCs as follows, and get the performance like in Table \ref{ECOC_LM}.

a. The polynomial CCs with k=2 and p=181. These Polynomial CCs are defined by
$
f: \mathbb{Z}/N\mathbb{Z} \longrightarrow \mathbb{F}_p^r 
$,
where $N=21901$, and $f_i(x)=((x \mod p)+floor(x/p)i) \mod p$, and r=2 or r=6. 

b. The Remainder CCs with k=2 and $p_i \in \{173,191, 157, 181, 193, 199\}$. These Remainder CCs are defined by
$
f: \mathbb{Z}/N\mathbb{Z} \longrightarrow \prod_{i=1}^r \mathbb{Z}/p_i\mathbb{Z} 
$, 
where $N=21901$, $f_i(x)=x \mod p_i$, and $r=2\mbox{ or }6$.

c. the Gauss CCs with k=2 and $p_i \in \{10 \pm9\sqrt{-1}, 13\pm2\sqrt{-1},  12\pm7\sqrt{-1}\}$. 
These Gauss CCs are defined by
$
f: \overline{U_{82}(0)} \cap \mathbb{Z}[\sqrt{-1}] \longrightarrow \prod _{i=1}^r 
\mathbb{Z}[\sqrt{-1}]/(p_i)
$, where $N=21901$, and $f_i(x)=x \mod (p_i)$, and r=2 or r=6.

d. ECOC of 15 bit.

\begin{table}[!hbp]
\tiny
\begin{tabular}{|c|c|c|c|c|c|c|c|}
\hline
\hline
ep.  & ECOC of 15 bit & Poly. CC of 2 sites &  Rem. CC of 2 sites & Gauss CC of 2 sites &  Poly. CC of 6 sites &  Rem. CC of 6 sites & Gauss CC of 6 sites\\
\hline
20	& 0.0069 &  0.0118 &	0.0081  & 0.0017 & 0.0640 & 0.0459 & 0.0230\\
 40 & 0.0795 & 0.6657 & 0.6130 & 0.4308 & 0.9878 &  0.9667& 0.9760	 \\
60	& 0.3660 &  0.8172 &  0.7629 & 0.8436 & 0.9968 & 0.9962 & 0.9966\\
80 & 0.5740 &  0.8684 &       0.8757 & 0.9195 & 0.9988 & 0.9983 & 0.9985\\
\hline
\hline
param. num ($10^7$) & $2.18 \times 15$ & $2.21\times 2$ & $2.21\times 2$& $2.21\times 2$ & $2.21 \times 6$ & $2.21 \times 6$ & $2.21 \times 6$\\
\hline
\hline
\end{tabular}
  \caption{Comparing of ECOC and CCs}
\label{ECOC_LM}
\end{table}

We can see, even when the base learner number 2 of CCs is much less than the base learner number 15 of ECOC,
the performance of CCs are better than the ECOC which trainable number of parameters of networks bigger than CCs.

\vspace{-2mm}
\section{Application for feature encode}
\vspace{-2mm}
For a categorical feature take value in $\mathbb{Z}/N\mathbb{Z}$, where $N$ is a 
huge integral number, we can use the composite mapping of a CC
$
    \mathbb{Z}/N\mathbb{Z}  \longrightarrow  \prod _{i=1} ^r \mathbb{Z}/N_i\mathbb{Z} 
$
and the nature embedding
\begin{displaymath}
  \begin{array}{ccl}
    \prod _{i=1} ^r \mathbb{Z}/N_i\mathbb{Z} & \longrightarrow & \prod_{i=1}^r \mathbb{F}_2^{N_i} = \mathbb{F}_2^{\sum_i N_i}\\
(x_i)_i & \mapsto &  (N_i \mbox{ bit one hot representation of } x_i )_i
  \end{array}
\end{displaymath}
to get a $r$-hot encoding. We use this $r$-hot encoding as feature encoding.

Apart from the CC feature encoding, the more natural ideas for feature encoding 
are
 
\textbf{COO}. Cut off of one-hot encoding. We call a $n$-bit binary code the 'Cut off of one-hot', if the $n-1$ most frequently used ID's are one-hot encoded in the
front $n-1$ bits, and all the other ID's are encoded to the code $'0\cdots 01'$.  

\textbf{RMP}. Using a code frequently used in error-correct encoding. For example, a 
Reed-Muller code \cite{RM} with punch by a random subset of bits. For a binary code $\{f_i\}_{i\in \mathbb{Z}/n\mathbb{Z}} : C \hookrightarrow \mathbb{F}_2^n$ and a subset $Q \subset \mathbb{Z}/n\mathbb{Z}$ 
 of $m$ elements,    
the punch of $f$ by $Q$ means the code $\{f_i\}_{i\in \mathbb{Z}/n\mathbb{Z} \setminus Q} : C \hookrightarrow 
\mathbb{F}_2^{n-m}$.

We will show that, the performance of our Polynomial CC, Remainder CC and Gauss 
CC are better than both the code COO and RMP.
\vspace{-2mm}
\subsection{Numeric Experiments}
\vspace{-2mm}

We use the dataset ``Movie Lends'' (\cite{Movie_Lends}), which has
the columns UserID, MovieID, Rating and Timestamp.
The UserIDs range between 1 and 6040, and MovieIDs range between 1 and 3952, ratings 
are made on a 5-star scale, timestamp is represented in seconds. 
Each user has at least 20 ratings. We use only the column UserID, 
MovieID and Rating. and use a DNN with  an embedding layer  and two full-connected layers.
In the embedding layer, the User code and Movie code are embedded 
to real vectors of dimension 32 respectively, the dimension of the output the two full-connected layers are
64 and 1 respectively. After the first full-connected layer we use 'RELU', after the second  full-connected
 layer we use $x\mapsto 4* sigmoid(x)+1$.  We use this network as a regression model, and train it by 
 minimize MSE. The ratio between train data and validation data 
 is 8:2.
We compare the validation loss of the following methods:

\textbf{1.} 582 bit cut off of the one-hot code for UserID, and 474 bit cut off of the one-hot 
code for MovieID.

\textbf{2.} 582 bit random punch of RM(12,1) for UserID, and 474 bit random punch of RM(11,1) 
for MovieID.

\textbf{3.} 582 bit 6-hot Polynomial code based on finite field $\mathbb{F}_{97}$ for UserID, and
474 bit 6-hot Polynomial code based on finite field $\mathbb{F}_{73}$ for MovieID.

\textbf{4.} 582 bit Remainder code with modules $\{83, 89, 97, 101, 103, 109\}$ for 
UserID, and 474 bit Remainder code with modules $\{67,71,73,79,83,101
\}$ for MovieID.

\textbf{5.} 582 bit Gauss code with modules $\{8 \pm 5\sqrt{-1}, 9 \pm 4\sqrt{-1}, 10+\sqrt{-1}, 10+3\sqrt{-1}\}$ for 
UserID, and 474 bit Remainder code with modules $\{67,71,73,79,83,101
\}$ for MovieID.

The validation losses are like in Table \ref{Coding_Methods}.
We see that the performance of Polynomial CC, Remainder CC and Gauss CC are 
better than the one-hot cut and RM code with punch of same length significantly. Moreover,
the performance of Gauss CC is best, and then the Remainder CC.

\vspace{-2mm}

\subsection{Theoretical analysis for feature coding}
\vspace{-2mm}

We see the performance of Polynomial CC, Remainder CC and Gauss CC are good for
feature coding, but we don't know how to choose the non-zero bit number $r$ in the coding. More 
generally, how to study the performance of codes without experiments? In the theory of error-correcting 
code, we know the Hamming distance is an important metric for codes. In general, if the original IDs
and length of coding is fixed, the error-correcting codes with big Hamming distance have good performance.
But for feature coding, Hamming distance is not a good metric. For example, we  compare
the performance of Method 2 introduced in the previous subsection and the anti-Method 2.  

The codings used in anti-Method 4 and Method 4 have the relationship: 
$x \mapsto 1-x$.
The corresponding pair of codes in the two method has same Hamming distance, but the performance is  
difference (in Table \ref{Coding_Methods}).
Hence the  Hamming distance is not a good choose for metric of feature 
encoding.

For a binary $r$-hot codeword $c$ of length $n$, we can view $\frac{1}{r}c$ as a distribution 
on $\mathbb{Z}/n\mathbb{Z}$,
 and call it \textbf{the reduced distribution of x}, write it as dist($c$). 
The \textbf{average minimal KL-divergence (AMKL)}  of a code $I \longrightarrow C$ is defined as
$
  \sum_i \min_j \mbox{KL}(dist(c_i) || dist(c_j))p_i
$.
We propose that use \textbf{AMKL} as the metric of code, 
and give the conjecture:
\begin{con}
  The feature code with bigger \textbf{AMKL} has better 
  performance.
  \label{con_KL}
\end{con}
To examine the conjecture \ref{con_KL}, we give a
lemma to compute the \textbf{AMKL} firstly:

\begin{lem}
  For a $n$ bit $r$-hot code $I \longrightarrow \mathbb{F}_2^n$,   
%
%
   if for any codeword $c_i$ the maximal common non-zero bit number between $c_i$ and any other codeword
  in $C$ is $r$, the \textbf{AMKL} equal to  
  $(1-\frac{t}{r})\infty$.
\end{lem}

\textbf{Proof.}
For any $i \in I$, let $x_i$ denote the codeword of $i$.
For any $i \neq j$ in $I$, the reduced distribution of $x_i, x_j$ are
  $\mbox{dist}(x)_i=\frac{1}{r}x_i$,  $\mbox{dist}(x)_j=\frac{1}{r}x_j$
respectively. 
Hence the KL-divergency of dist($x_i$), dist($x_j$) is
$\mbox{KL}(dist(x_i)||dist(x_j)) 
=  \frac{1}{r}\log\frac{1/r}{0} \times (r-\tau)+ \frac{1}{r}\log\frac{1/r}{1/r} 
  \times \tau  
  = (1-\frac{\tau}{r})\log\infty$.
Hence $\sum_i \min_j \mbox{KL}(dist(c_i) || dist(c_j))p_i=
\sum_i (1-\frac{\tau}{r})p_i\log\infty=(1-\frac{\tau}{r})\log\infty$.
%
\qed

We use the some numeric experiments to examine the conjecture \ref{con_KL}.
We use the following encoding Method on dataset ``Movie Lends'', 
 and their \textbf{AMKL} and performance is like 
in table \ref{avg_min_KL}. We see that the \textbf{AMKL} has 
positive effect to performance. 
Moreover, the performance of Gauss CC > Remainder CC > Polynomial CC with same length and 
\textbf{AMKL}.

\textbf{Method 1.} 582 bit Remainder code with modules $\{289,293\}$ for UserIDs, and 474 bit 
Remainder code with modules $\{235, 239\}$ for MovieIDs.

\textbf{Method 2.} 582 bit Remainder code with modules $\{193, 194, 195\}$ for UserIDs, and 474 bit 
Remainder code with modules $\{157, 158, 159\}$ for MovieIDs.

\textbf{Method 3.} 582 bit 6-hot Polynomial code based on finite field $\mathbb{F}_{97}$ for UserIDs, and
474 bit 6-hot Polynomial code based on finite field $\mathbb{F}_{73}$ for MovieIDs.

\textbf{Method 4.} 582 bit Remainder code with modules $\{83, 89, 97, 101, 103, 109\}$ for 
UserIDs, and 474 bit Remainder code with modules $\{67,71,73,79,83,101
\}$ for MovieIDs.

\textbf{Method 5.} 582 bit Gauss code with modules $\{8 \pm 5\sqrt{-1}, 9 \pm 4\sqrt{-1}, 10+\sqrt{-1}, 10+3\sqrt{-1}\}$ for 
UserIDs, and 474 bit Remainder code with modules $\{67,71,73,79,83,101
\}$ for MovieIDs.

\textbf{Method 6.} 582 bit Remainder code with modules \{19, 23, 25, 27, 29, 31, 32, 37, 41, 43, 47, 49, 53, 59, 67
\}
for UserIDs, and 473 bit 
Remainder code with modules \{17, 19, 23, 25, 27, 29, 31, 32, 37, 41, 43, 47, 49, 53\} 
for MovieIDs.


\begin{table}[!hbp]
    \tiny
 \begin{minipage}{0.6\linewidth}
\begin{tabular}{|c|c|c|c|c|c|c|}
\hline
\hline
ep.  & Method 1 & Method 2 &  Method 3 & Method 4 & Method 5 & anti-Meth. 4\\
\hline
1 &  1.044   & 0.960 & 0.938  &  0.936  &  0.932 & 1.235 \\
\hline
5 & 1.036  & 0.862 & 0.857  &  0.857  &   0.854 &  1.301\\
\hline
9 &  1.034 & 0.857 & 0.853  &  0.851  &  0.849 & 1.177 \\
\hline
13 & 1.029 & 0.855 &  0.853  & 0.850  &   0.846 & 1.141 \\
\hline
\hline
\end{tabular}
  \caption{Comparing of Coding Methods}
\label{Coding_Methods}
 \end{minipage}
 \begin{minipage}{0.5\linewidth}
   \begin{tabular}{|c|c|c|c|c|c|}
\hline
\hline
method  &  u. avg min KL &  i. avg min KL & MSE ep.1 & MSE ep. 2 & MSE ep.3 \\
\hline
1 (two hot) & 0.5 & 0.5 & 1.0335 & 0.951 & 0.913 \\
\hline
2 (three hot) & 0.667 & 0.667 & 0.992 & 0.912 & 0.887\\
\hline
3 (six hot) & 0.833 & 0.833 & 0.938 & 0.886 & 0.873\\
\hline
4 (six hot) & 0.833 & 0.833 & 0.936 & 0.883 & 0.867 \\
\hline
5 (six hot) & 0.833 & 0.833 & 0.932 & 0.883 & 0.866 \\
\hline
6 (15, 14 hot) & 0.867 & 0.857 & 0.908 & 0.875 & 0.864\\
\hline
\hline
\end{tabular}
  \caption{avg min KL}
\label{avg_min_KL}
 \end{minipage}
\end{table} 

\vspace{-2mm}
\section{Conclusion}
\vspace{-2mm}
We propose three classes of category coding (CC) with minimal collision property. 
They are Polynomial CC, Remainder CC and Gauss CC.

In 
the application for label coding in the classification problem with huge labels number using CNN, 
we prove that they have good theoretical properties and show that they have good
performance in numerical experiments. 

In the application for feature coding in 
collaborative filtering using DNN, we show that their performance is better than cut-off method
and  classical binary coding method. Moreover, we give a metric ``AMKL''of feature 
coding, and show it has positive effect to the performance. In additional,
we show that the performance of Gauss CC > Remainder CC > Polynomial CC with same length and 
AMKL.

%
%
%
%
%
%
%
%
%
%
%
%
%
%
%
%
%
%
%
%
%
%
%
%
%
%
%
%
%
%
%

%

\end{document}